\documentclass[intlimits,twoside,a4paper]{article}

\usepackage[cp1251]{inputenc}

\usepackage[eqsecnum]{cmpj3}

\issue{2018}{21}{1}{13601}
\doinumber{10.5488/CMP.21.13601}

\title[First-principles study of properties of Al$_{3}$Ta compound]{First-principles study of the structural, phonon, elastic, and thermodynamic properties of Al$_{3}$Ta compound under high pressure}
\author[W. Leini, T. Zhang, Z. Wu, N. Wei]{W. Leini\refaddr{label1},
        T. Zhang\refaddr{label2}, Z. Wu\refaddr{label3}, N. Wei\refaddr{label3}}
\addresses{
\addr{label1} Anhui Sanlian University, Hefei 230601, P. R. China
\addr{label2} Chinese Academy of Sciences, Hefei 230031, P. R. China
\addr{label3} Anhui University, Hefei 230601, P. R. China
}

\date{Received August 23, 2017, in final form October 9, 2017}

\begin{document}
\maketitle
\begin{abstract}
We have investigated the phonon, elastic and thermodynamic properties of L1$_{2}$ phase Al$_{3}$Ta by density functional theory approach combining with quasi-harmonic approximation model. The results of phonon band structure shows that L1$_{2}$ phase Al$_{3}$Ta possesses dynamical stability in the pressure range from 0 to 80~GPa due to the absence of imaginary frequencies. The pressure dependences of the elastic constants $C_{ij}$, bulk modulus~$B$, shear modulus~$G$, Young's modulus $Y$, $B/G$ and Poisson's ratio $\nu$ have been analysed. The elastic constants are satisfied with mechanical stability criteria up to the external pressure of 80~GPa. The results of the elastic properties studies show that Al$_{3}$Ta compound possesses a higher hardness, improved ductility and plasticity under higher pressures. Further, we systematically investigate
the thermodynamic properties, such as the Debye temperature $\Theta$, heat capacity $C_{p}$, and thermal expansion coefficient $\alpha$, and provide the relationships between thermal parameters and pressure.
\pacs 61.82.Bg, 62.20.dc, 71.20.Be, 71.15.Mb
\end{abstract}

\section{Introduction}

 The L1$_{2}$ type trialuminide compounds Al$_{3}$M (where ``M'' represents transition or rare earth elements) have increasingly attracted attention due to the outstanding mechanical properties such as high specific strength and elastic moduli \cite{Milman,ZhangY,Jahnatek,Duan}. Moreover, they also possess low density, high melting points, superior oxidation resistance, sufficient creep resistance, good thermal stability and conductivity \cite{Harada,Morrison,Knipling,Chang}. All these excellent properties enable them as the ideal dispersed strengthening phases for the high-strength thermally-stable Al based alloys. However, the poor creep-resistance properties and the lack of ductility have hindered their industrial applications with the elevated temperatures \cite{Yamaguchi,Sasaki,Asta}. Refractory metals are alloyed for the purpose of improvement of low temperature ductility, creep resistance, oxidation resistance and toughness. The most widely used metallic elements are tungsten, molybdenum, rhodium, tantalum and niobium. In these refractory metals, tantalum is resistant to corrosion from acids, organic chemicals and aqueous solutions of salts. Tantalum is also used to produce a variety of alloys that have high melting points, strength and ductility. 

There are only a few theoretical and experimental studies dealing with the structural, phonon, elastic, and electronic properties of Al$_{3}$Ta in the literature. Asta et al. proposed that Al$_{3}$Ta is  considered to be a potential electronic material for very large scale integration applications because the compound reduces the low temperature interdiffusion barrier between aluminium and silicon \cite{Asta}. Al$_{3}$Ta is among trialuminide compounds  characterized by a melting point of 1823~K and a density of 6.9~g/cm. Boulechfar et al. investigated the phase stability and electronic properties in Al$_{3}$Ta compound using the FP-LAPW method \cite{Boulechfar}. They found that there is observed a characteristic of covalent bonding in Al$_{3}$Ta compound. However, it is not clear whether the cubic L1$_{2}$ phase is the phase observed experimentally. Some researchers suggest that  L1$_{2}$ phase Al$_{3}$Ta is relatively stable \cite{Carlsson,Colinet}. To our best knowledge, no systematic experimental and theoretical investigations on the thermoelasticity for L1$_{2}$ Al$_{3}$Ta are performed up till now, which could not reflect the high temperature mechanical and elastic behaviour. 

Temperature dependent elastic properties are crucial for high temperature applications of alloys. Moreover, knowledge of thermoelasticity is also essential for predicting and understanding material response, strength, mechanical stability as well as phase transitions under high temperature \cite{ChunyangZhao}. High pressure and temperature can cause large effects on chemical and physical properties of a solid. As we know, the fact that high pressure is not easy to reach and control in experiment condition, while adjusting pressure in theoretical simulations can be accomplished straightly by changing the size of a unit cell.  For elastic constants, they offer a link between the elastic and dynamic behaviours of solid materials and provide important information on the nature of forces in a material \cite{HuYan,Jafari}. We can obtain the elastic parameters such as bulk modulus $B$, shear modulus $G$, Young's modulus $Y$, $B/G$ and Poisson's ratio $\nu$ by elastic constants. Moreover, the thermodynamic properties under high pressure and temperature are of great interest to geophysicists and physicists. To get a better understanding of thermodynamic properties of L1$_{2}$ phase Al$_{3}$Ta, more temperature-dependent parameters such as the Debye temperature $\Theta$, specific heat $C_{p}$, and thermal expansion coefficient $\alpha$ are required. In this paper, we firstly focus on investigating the stability of L1$_{2}$ phase Al$_{3}$Ta through the lattice dynamics study. Then, we investigate elastic and thermodynamic properties of Al$_{3}$Ta under high pressure. We believe this work can help us in designing and understanding the high pressure behaviour of Al$_{3}$Ta.

\section{Methods}

In the present work, all the calculations were performed by using first-principles based on the plane wave pseudopotential density-function theory (DFT) method, which are carried out on the Quantum ESPRESSO code \cite{Baroni,Giannozzi}. We calculate the elastic constants by ElaStic tool which can be interfaced with computer packages WIEN2k and Quantum ESPRESSO \cite{Golesorkhtabar,Blanco}. We have used the generalized gradient approximation (GGA) parameterized by Perdew-Burke-Ernzerhof (PBE) for the exchange and correlation terms in the electron-electron interaction for $k$-space summation which was $12\times12\times12$ Monkhorst and Pack grid of $k$-points. The kinetic energy cutoff for wavefunctions is 50~Ry, and the kinetic energy cutoff for charge density and potential is 500~Ry. The convergence threshold for selfconsistency is $10^{-8}$~Ry. 

In order to investigate the thermodynamic properties of Al$_{3}$Ta, we use the quasi-harmonic Debye model implemented in the Gibbs program \cite{Blanco}. The key procedure for thermoelastic calculations is to compute the second derivatives of non-equilibrium Gibbs function $G^{*}(V;P,T)$ with respect to the applied strain. For a given volume $V$ and temperature $T$, non-equilibrium Gibbs function $G^{*}(V;P,T)$ can be written as:
\begin{equation}\label{(1)}
   {G^*}(V;P,T) = E(V) + PV + {A_{\text{vib}}}(\Theta ,T), 
\end{equation}
where $E(V)$ is total energy per unit cell of Al$_{3}$Ta, $P$ is the hydrostatic pressure, $A_{\text{vib}}(\Theta,T)$ represents the vibrational Helmholtz free energy which can be taken as:
\begin{equation}\label{(2)}
   {A_{\text{vib}}}(\Theta ,T) = nKT\left[ {\frac{{9\Theta }}{{8T}} + 3\ln \left(1 - {\re^{ - \frac{\Theta }{T}}}\right) - D\left(\frac{\Theta }{T}\right)} \right],  
\end{equation}
where $D(\Theta/T)$ is the Debye integral, and $n$ is the number of atoms per formula unit, $\Theta$ takes the form of:
\begin{equation}\label{(3)}
   \Theta  = \frac{\hbar }{K}{\left(6{\piup ^2}{V^{\frac{1}{2}}}n\right)^{\frac{1}{3}}}f(\nu )\sqrt {\frac{{{B_S}}}{M}}\,,
\end{equation}
where $B_{S}$ is the adiabatic bulk modulus, $M$ is the molecular mass per formula unit, which can be expressed in the form: 
\begin{equation}\label{(4)}
   {B_S} = V\frac{{{\rd^2}E(V)}}{{\rd{V^2}}}.  
\end{equation}
And $f(\nu)$ is given by:
\begin{equation}\label{(5)}
   f(\nu ) = {\left\{ {3{{\left[ {2{{\left(\frac{{21 + \nu }}{{31 - \nu }}\right)}^{3/2}} + {{\left(\frac{1}{3}\frac{{1 + \nu }}{{1 - \nu }}\right)}^{3/2}}} \right]}^{ - 1}}} \right\}^{1/3},} 
\end{equation}
where $\nu$ is Poisson's ratio. Hence, the  non-equilibrium Gibbs function $G^{*}(V;P,T)$ as a function of $(V;P,T)$ can be minimized with respect to the volume as
\begin{equation}\label{(6)}
   {\left( {\frac{{\rd{G^*}(V;P,T)}}{{\rd V}}} \right)_{P,T}} = 0.
\end{equation}
By solving equation~(\ref{(6)}), one can get the thermal equation of state $V(P,T)$. After the equilibrium state of a given $V(P,T)$ has been gained, the isothermal bulk modulus and other thermodynamic properties, such as the heat capacity, vibrational internal energy, and thermal expansion $\alpha$ can be evaluated in the appropriate thermodynamic expressions.
\begin{equation}\label{(7)}
   {C_V} = 3nK\left[ {4D\left(\frac{\Theta }{T}\right) - \frac{{3\Theta /T}}{{{\re^{\Theta /T}} - 1}}} \right],
\end{equation}
\begin{equation}\label{(8)}
   {C_P} = 3nK\left[ {4D\left(\frac{\Theta }{T}\right) - \frac{{3\Theta /T}}{{{\re^{\Theta /T}} - 1}}} \right](1 + \alpha \gamma T), 
\end{equation}
\begin{equation}\label{(9)}
   \alpha  = \frac{{\gamma {C_V}}}{{BV}}\,,
\end{equation}
where $\gamma$ is the Gr\"{u}neisen parameter. This method has already been successfully used to investigate the thermodynamic properties of a series of compounds \cite{ZhangXY2,Sun,Florez,Chen}.

\section{Results and discussions}
\subsection{Structural and phonon properties}
\begin{figure}[!t]
\centering
\includegraphics[scale=0.65]{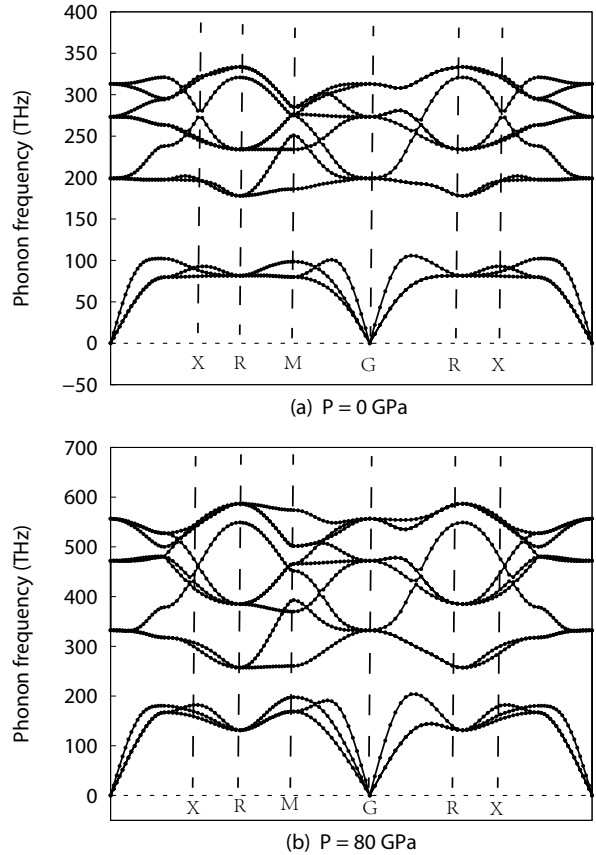}
\caption{The phonon spectra of L1$_{2}$ phase Al$_{3}$Ta at pressure of 0 and 80~GPa, respectively.}
\label{Fig1}
\end{figure}

Firstly, the equilibrium lattice parameters have been computed by minimizing the crystal total energy calculated for different values of a lattice constant. We calculated the ground-state lattice parameters of L1$_{2}$ phase Al$_{3}$Ta (space group: Pm$\bar{3}$m, No:~221) alloys at 0~GPa. As expected, the optimized lattice parameter of Al$_{3}$Ta obtained from GGA is found to be 4.024~{\AA}, there are no experimental results to compare with, while it is in agreement with the other theoretical value of 4.018~{\AA} \cite{Boulechfar}. This agreement provides a confirmation such that the computational methodology utilized in our work is suitable  and reliable. In order to gain an insight into the phase stability of L1$_{2}$ phase Al$_{3}$Ta under high pressure, the phonon band structure has been studied. The phonon band structure of L1$_{2}$ phase Al$_{3}$Ta along some high symmetry directions in the Brillouin zone at 0~GPa was displayed  in figure~\ref{Fig1}~(a). These curves are very similar to those obtained for other platinum-based alloys in the same structure. The calculated phonon dispersion curves do not contain a soft mode at any vectors, which confirms the stability of L1$_{2}$ phase Al$_{3}$Ta at 0~GPa. The unit cell of Al$_{3}$Ta has four atoms, which give rise to a total of 12 phonon branches, which contains three acoustic modes and nine optical modes. We also provide the phonon curves of L1$_{2}$ phase Al$_{3}$Ta at 80~GPa in figure~\ref{Fig1}~(b). It is obvious that Al$_{3}$Ta is stable under pressure of 80~GPa due to the absence of imaginary frequencies. Moreover, we can see the phonon band structure at 80~GPa shift to the high energy on the whole comparing with that at 0~GPa, meaning that the frequencies of the phonon increase as the pressure increases. Those results indicate that  L1$_{2}$ phase Al$_{3}$Ta possesses dynamical stability in the pressure range from 0 to 80~GPa and ensure the subsequent study being credible.  

\subsection{Mechanical properties}
\begin{figure}[!t]
\centering
\includegraphics[scale=0.65]{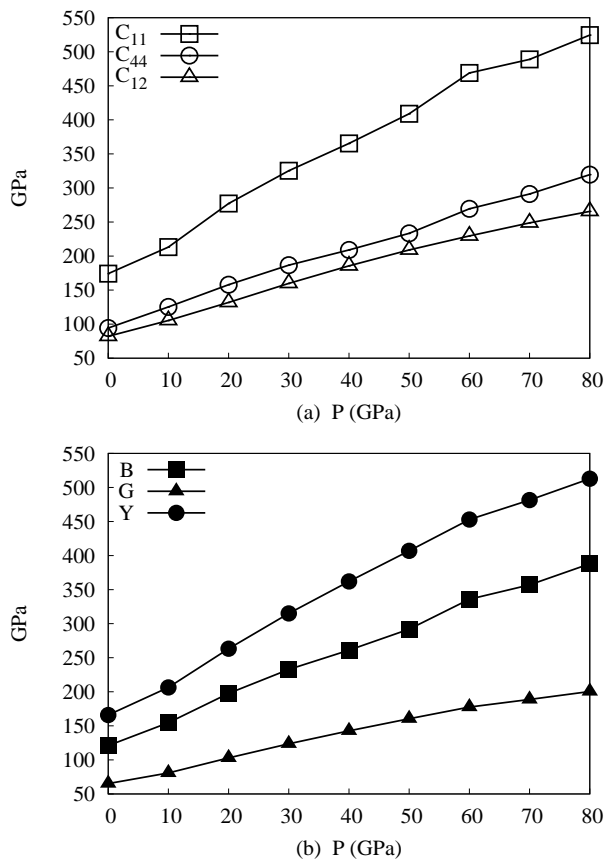}
\caption{The calculated elastic constants $C_{ij}$ (a), the bulk modulus $B$, shear modulus $G$ and Young's modulus $Y$ (b) as a function of pressure for L1$_{2}$ phase Al$_{3}$Ta.}
\label{Fig2}
\end{figure}

 Elastic properties of a solid can provide important information on the mechanical, dynamic and thermodynamic behaviours of materials. In this work, firstly, we predict the elastic constants of Al$_{3}$Ta in the pressure range of 0--80~GPa at 0~K. The elastic constants of solids provide important information concerning the nature of the forces operating in solids and provide information about the stability and stiffness of materials. The elastic constants of a material can be obtained using the stress-strain method by calculating the total energy as a function of lattice deformation. In the method, a small strain should be loaded on a crystal. The elastic constants are defined by means of a Taylor expansion of the total energy~$E(V)$ with respect to a small strain \cite{ZhangXY}. The energy $E(V)$ is given as follows:
\begin{equation}\label{(10)}
   E(V) = E({V_0}\,,0) + \frac{1}{2}\sum\limits_i^6 {\sum\limits_j^6 {{C_{ij}}{\varepsilon _i}{\varepsilon _j}} }\,,
\end{equation} 
where $V_{0}$ is the volume at ground state; $C_{ij}$ represents elastic constants, $\varepsilon_{i}$ and $\varepsilon_{j}$ represent the strain.
For L1$_{2}$ phase Al$_{3}$Ta, there are three different independent elastic constants, $C_{11}$, $C_{12}$, and $C_{44}$. At 0~GPa, the calculated values of elastic constants $C_{11}=174.18$~GPa, $C_{12}=94.30$~GPa, and $C_{44}=82.15$~GPa. There is no doubt that the elastic constants of a material are strongly affected by pressure. Figure~\ref{Fig2}~(a) shows the pressure dependences of elastic constants  $C_{ij}$. We can analyse the mechanical stability of L1$_{2}$ phase Al$_{3}$Ta based on the elastic constants. The traditional mechanical stability conditions in cubic crystals on the elastic constants are known as:  $C_{11}> 0$, $C_{12}> 0$, $C_{11}-C_{12} > 0$, and  $C_{11}+2C_{12} > 0$ \cite{Lin,Abraham}. It is obvious that all the elastic constants of Al$_{3}$Ta in a wide pressure range (0--80~GPa) satisfy these traditional stability conditions, meaning that L1$_{2}$ phase Al$_{3}$Ta is mechanically stable under pressure up to 80~GPa. This result is in agreement with the phonon band structure discussions and results. From figure~\ref{Fig2}~(a), all elastic constants of Al$_{3}$Ta increase almost monotonously with an increase of pressure and $C_{11}$ has slightly smaller amplitude. This is attributed to the lattice parameters becoming lower under high pressure.

It is acknowledged that based on the elastic constants, the polycrystalline bulk modulus $B$, shear modulus $G$ and Young's modulus $Y$ can be estimated based on the Voigte-Reusse-Hill approximation~\cite{XinLi}. In general, considering the fundamental physics of a solid, the bulk modulus $B$ is usually assumed to be a fundamental physical property of solids and is used as a measure of the average bond strength of the atoms for particular crystals. At the same time, the shear modulus $G$ is a measure of resistance to reversible deformations upon shear stress while the large value of shear modulus $G$ is an indication of a more pronounced directional bonding between atoms. Young's modulus $Y$ is defined as the ratio of the tensile stress to the corresponding tensile strain, and is an important quantity for technological and engineering applications. We display a relationship between the bulk modulus $B$, shear modulus~$G$ and Young's modulus $Y$ with the pressure in figure~\ref{Fig2}~(b). The value of bulk  modulus $B$ and shear modulus $G$ is 120.13~GPa and 65.26~GPa, respectively at 0~GPa. This result is in good agreement with Boulechfar et al. results \cite{Boulechfar}. It is seen that for Al$_{3}$Ta, bulk modulus $B$, shear modulus~$G$ and Young's modulus $Y$ increase almost linearly with an increasing pressure, where $B$ and $Y$ have almost the same variation amplitudes, and $G$ has a slightly smaller amplitude, indicating that the effect of pressure on these quantities are prominent.  This is attributed to the atoms distance in the interlayers becoming shorter, and the interactions between these atoms becoming stronger. 

\begin{figure}[!t]
\centering
\includegraphics[scale=0.65]{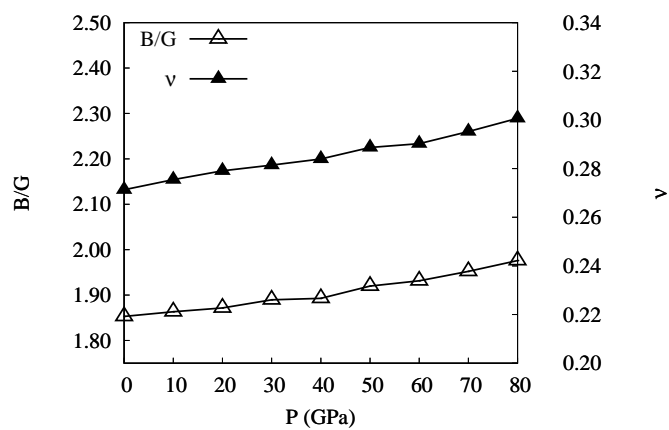}
%\vspace{-3mm}
\caption{The value of $B/G$ and Poisson's ratio $\nu$ as a function of pressure for L1$_{2}$ phase Al$_{3}$Ta.}
\label{Fig3}
\end{figure}

To further analyse the mechanical behaviour of L1$_{2}$ phase Al$_{3}$Ta, the ductile or brittle behaviour should be discussed. As we know, the bulk modulus $B$ and shear modulus $G$ represent a resistance to plastic deformation and a resistance to fracture, respectively. According to Pugh formation, $B/G$ ratio has been proposed to predict a brittle or ductile behaviour \cite{Pugh}. A high $B/G$ value is associated with ductility, while a low $B/G$ value corresponds to brittleness. If $B/G < 1.75$, a material exhibits a brittle behaviour. Otherwise, it exhibits a ductile behaviour \cite{Jossou,Zheng,Ivashchenko}. Figure~\ref{Fig3} presents the calculated $B/G$ value of L1$_{2}$ phase Al$_{3}$Ta as a function of pressure. We observe that $B/G$ value increases from 1.85 to 1.99 when the pressure increases from 0 to 80~GPa. Those suggest that Al$_{3}$Ta exhibits a ductility characteristic, and thus its ductility increased under high pressure. Poisson's ratio $\nu$ is consistent with $B/G$ ratio, which is also related to the brittleness. The critical value which separates ductile and brittle material is 0.26. If Poisson's ratio $\nu> 0.26$, the material behaves in a ductile manner. Otherwise, the material behaves in a brittle manner. In addition, Poisson's ratio is usually used to quantify the stability of the crystal against shear. The larger is the Poisson's ratio, the better is the plasticity. We note that the value of Poisson's ratio $\nu$ increases from 0.274 to 0.302 as the pressure increases from 0 to 80~GPa  in figure~\ref{Fig3}, suggesting that L1$_{2}$ phase Al$_{3}$Ta exhibits a ductility under high pressure and the pressure can improve the ductility while the existence of an external pressure can improve the plasticity.  

\subsection{Thermodynamic properties}
\begin{figure}[!t]
\centering
\includegraphics[width=\textwidth]{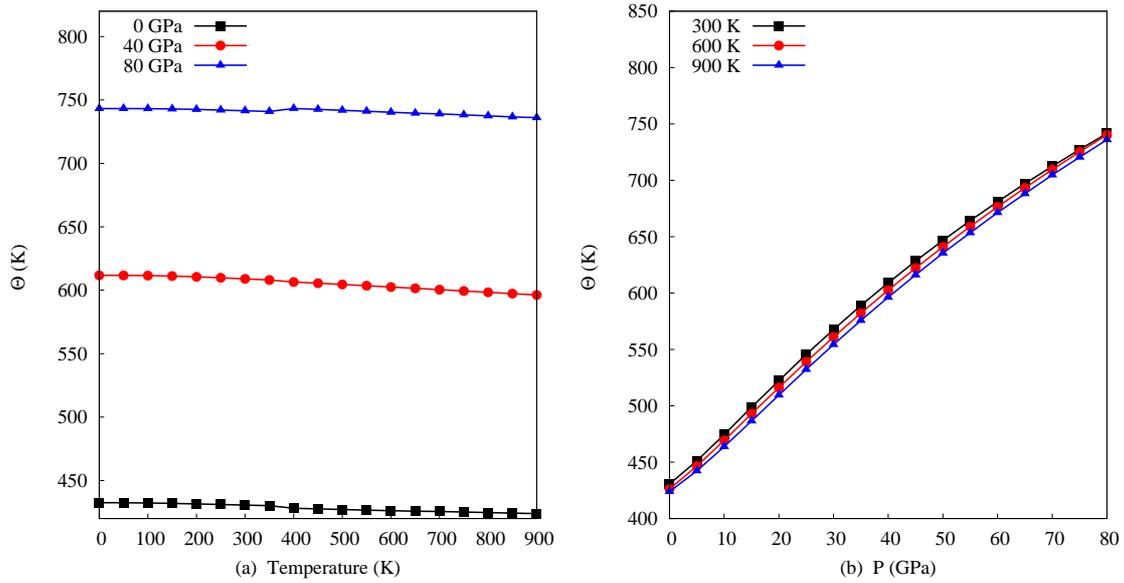}
\caption{(Colour online) The Debye temperature $\Theta$ as a function of temperature at different pressures of $P=0$, 40, and 80~GPa for L1$_{2}$ phase Al$_{3}$Ta~(a). The Debye temperature $\Theta$ as a function of pressure at different temperatures of $T=300$, 600, and 900~K for L1$_{2}$ phase Al$_{3}$Ta~(b).}
\label{Fig4}
\end{figure}

The data of thermodynamic properties under high pressure and temperature can provide a valuable information for industrial applications of materials under extreme conditions. We employ the quasi-harmonic Debye model to obtain the thermodynamic properties of L1$_{2}$ phase Al$_{3}$Ta at various temperatures  and pressures from the energy-volume relation. As one of the important physical quantities for a solid, the Debye temperature $\Theta$  is a an important parameter describing the material thermodynamic properties in solid state physics. It is closely related to specific heat, bond strength, elastic stiffness constants and melting temperature. Figure~\ref{Fig4} displays the dependence of the Debye temperature of Al$_{3}$Ta on temperature and pressure. Variations in Debye temperature $\Theta$ versus temperature at different fixed pressures, which are $P=0, 40$, and 80~GPa, are shown in figure~\ref{Fig4}~(a). From the figure, it is obvious that Debye temperature~$\Theta$ in the range of temperature from 0~K to 900~K remains approximately unaltered, suggesting that Debye temperature~$\Theta$ is insensitive to the temperature. In figure~\ref{Fig4}~(b), the Debye temperature $\Theta$ increases almost linearly with applied pressures at a given temperature, which indicates the change of the vibration frequency of atoms under pressure.  This may be attributed to the influence of the isothermal bulk modulus which is independent of pressure. Hence, the pressure has a more significant effect on the Debye temperature $\Theta$ of L1$_{2}$ phase Al$_{3}$Ta  comparing with the temperature. These results are consistent with the general behaviour of a decreasing Debye temperature with an increase of temperature in other L1$_{2}$ structures \cite{Surucu,Surucu2}. 

\begin{figure}[!t]
\centering
\includegraphics[width=\textwidth]{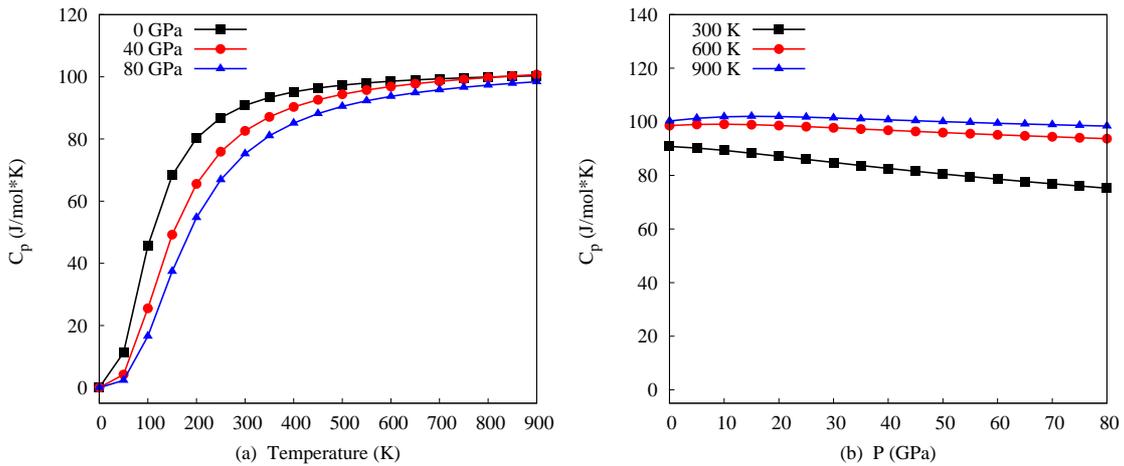}
\caption{(Colour online) The heat capacity $C_{p}$  as a function of temperature at different pressures of $P=0$, 40, and 80~GPa for L1$_{2}$ phase Al$_{3}$Ta~(a). The heat capacity $C_{p}$ as a function of pressure at different temperatures of $T=300$, 600, and 900~K for L1$_{2}$ phase Al$_{3}$Ta~(b).
}
\label{Fig5}
\end{figure}

The heat capacity is an important parameter of a material. Knowledge of the heat capacity of a solid not only provides an essential insight into its vibrational properties but also offers some instructions for many applications. In figure~\ref{Fig5}~(a), we show the heat capacity $C_{p}$ as a function of temperature  $T$ at the pressures of 0, 40, and 80~GPa.  It is realized from the figure that when $T < 300$~K, the $C_{p}$ increases very rapidly with the temperature; when $T > 300$~K, the heat capacity $C_{p}$ increases slowly with the temperature, and it almost approaches a constant value referred to as Dulong-Petit limit for this compound. Additionally, the heat capacity $C_{p}$ rapidly approaches zero while the temperature approaches  absolute zero. In figure~\ref{Fig5}~(b), we depict the variation of heat capacity with pressure at a fixed temperature of 300~K, 600~K, and 900~K. It is obvious that the heat capacity keeps almost unchanged with an increase of the pressure  and the further compression slows down this trend. This reveals that the heat capacity~$C_{p}$ is mainly dependent on the temperature, and a higher temperature almost does not affect the heat capacity~$C_{p}$.

\begin{figure}[!t]
\centering
\includegraphics[width=\textwidth]{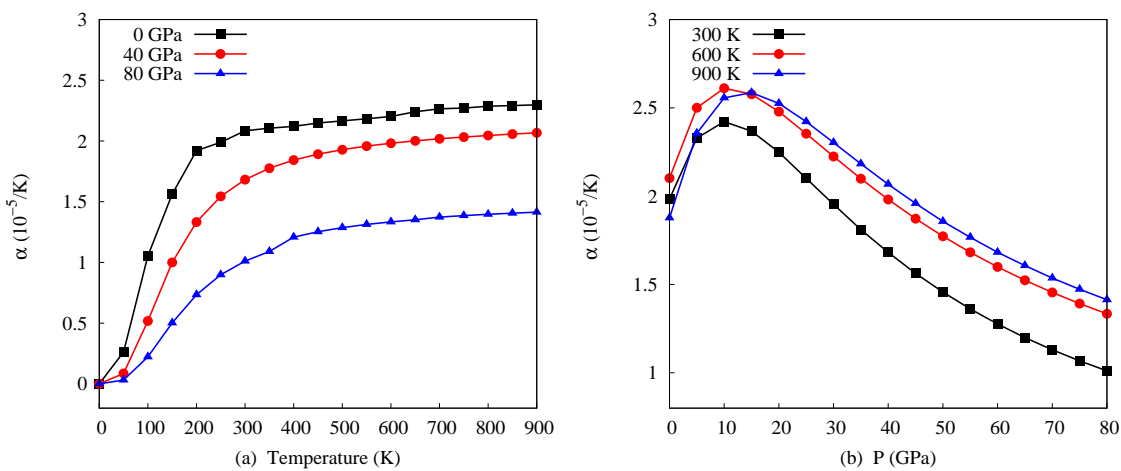}
\caption{(Colour online) The volume thermal expansion coefficient $\alpha$ as a function of temperature at different pressures of $P=0$, 40, and 80~GPa for L1$_{2}$ phase Al$_{3}$Ta~(a). The volume thermal expansion coefficient $\alpha$ as a function of pressure at different temperatures of $T=300$, 600, and 900~K for L1$_{2}$ phase Al$_{3}$Ta~(b).
}
\label{Fig6}
\end{figure}

The variation of the thermal expansion coefficient $\alpha$ of L1$_{2}$ phase Al$_{3}$Ta with temperature and various pressures is shown in figure~\ref{Fig6}. It is clearly seen that $\alpha$ exhibits a similar trend for all isobars, and at 0~GPa, the thermal expansion coefficient $\alpha$ increases exponentially with $T$ in the low temperature region and becomes flat at high temperatures in figure~\ref{Fig6}~(a). The thermal expansion coefficient possesses the highest values for lowest ($P = 0$~GPa) pressure in all the temperature range considered. Moreover, as seen from figure~\ref{Fig6}~(b), the thermal expansion coefficient $\alpha$ increases with an increasing pressure at a given temperature under 10~GPa and then decreases with an increasing pressure. This suggests us that L1$_{2}$ phase Al$_{3}$Ta will have the largest thermal expansion coefficient $\alpha$ value at about 10~GPa.

\section{Conclusions }
In this work, the structural, phonon, elastic and thermodynamic properties of Al$_{3}$Ta compound under high pressures were theoretically investigated by performing GGA calculations based on DFT method. The phonon band structure of L1$_{2}$ phase Al$_{3}$Ta indicates that it possesses a dynamical stability in the whole studied pressure range from 0 to 80~GPa due to the absence of imaginary frequencies. The pressure dependence of the elastic constants C$_{ij}$, bulk modulus $B$, shear modulus $G$, Young's modulus $Y$, $B/G$, and Poisson's ratio $\nu$ are successfully calculated. The results of the elastic properties studies show that Al$_{3}$Ta compound is mechanically stable and possesses a higher hardness, improved ductility and plasticity under higher pressures. Moreover, we have studied the thermodynamic properties, such as the Debye temperature $\Theta$, heat capacity $C_{p}$, and thermal expansion coefficient $\alpha$ using the quasi-harmonic Debye model in the range of temperatures from 0~K to 900~K.

\section*{Acknowledgements}
The calculations were performed in Center for Computational Science on the ScGrid of Supercomputing Center, Computer Network Information Center of Anhui university and Chinese Academy of Sciences. This work was supported by the National Science Foundation of China under Grants Nos.~11774284 and the Anhui Provincial Natural Science Foundation of China (1508085SME219).

\newpage
\ukrainianpart
\title{Першопринципні дослідження  структурних, фононних, пружних та термодинамічних 
властивостей сполуки Al$_{3}$Ta при високому тиску}
\author{В. Леіні\refaddr{label1}, Т. Шенг\refaddr{label2}, Ш. Ву\refaddr{label3}, Н. Вей\refaddr{label3}}
\addresses{
\addr{label1} Аньхойський університет Саньлянь, Хефей  230601, Китай
\addr{label2} Китайська академія наук, Хефей 230031, Китай
\addr{label3} Аньхойський університет, Хефей  230601, Китай
}

\makeukrtitle
\begin{abstract}
Досліджено фононні, пружні та термодинамічні властивості  Al$_{3}$Ta  у  L1$_{2}$ фазі  методом функціоналу густини 
у поєднанні з квазигармонічною апроксимаційною моделлю. Результати структури фононної зони  показують, що 
 Al$_{3}$Ta  у L1$_{2}$ фазі  володіє динамічною стійкістю  у діапазоні тиску від 0 до 80~ГПа 
 завдяки відсутності  уявних частот. Проаналізовано тискові залежності  $C_{ij}$, об'ємного модуля пружності $B$, 
 модуля зсуву $G$, модуля Янга $Y$, $B/G$ та коефіцієнта Пуассона $\nu$. 
 Пружні сталі задовольняють критерій механічної стійкості  аж до зовнішнього тиску  80~ГПа. 
 Результати аналізу властивостей пружності  показують, що сполука  Al$_{3}$Ta 
 володіє вищою твердістю, кращою тягучістю і пластичністю  при високих тисках.
 Крім того, систематично досліджено такі  термодинамічні властивості,  як температура Дебая   $\Theta$, питома теплоємність 
 $C_{p}$ та коефіцієнт теплового розширення  $\alpha$, а також встановлено залежність між тепловими параметрами і тиском.
\keywords перші принципи, фонон, пружні властивості, термодинамічні властивості
\end{abstract}

\end{document}